\documentclass[a4paper,11pt]{article}
\usepackage{jheppub}
\usepackage[T1]{fontenc} 
\title{\boldmath Quantum unitary dynamics of a charged fermionic field and Schwinger effect}

\author[a]{\'Alvaro \'Alvarez-Dom\'inguez,}
\author[a,b]{Luis J. Garay,}
\author[c,d]{David Garc\'ia-Heredia,}
\author[a]{Mercedes Mart\'in-Benito.}

\affiliation[a]{Departamento de F\'isica Te\'orica and IPARCOS, Universidad Complutense de Madrid, Plaza de las Ciencias 1, 28040 Madrid, Spain}
\affiliation[b]{Instituto de Estructura de la Materia (IEM-CSIC), Serrano 121, 28006 Madrid, Spain}
\affiliation[c]{Ludwig-Maximilans Universit\"at, Geschwister-Scholl-Platz 1, 80539 Munich, Germany}
\affiliation[d]{Technische Universit\"at M\"unchen, Arcisstra\ss e 21, 80333 Munich, Germany}

\emailAdd{alvalv04@ucm.es}
\emailAdd{luisj.garay@ucm.es}
\emailAdd{David.Garcia@campus.lmu.de}
\emailAdd{m.martin.benito@ucm.es}

\usepackage{amssymb}
\usepackage[T1]{fontenc}
\usepackage[utf8]{inputenc}
\usepackage{physics}
\usepackage{amsmath,amssymb,amsthm,textcomp}
\usepackage{mathrsfs}
\usepackage{bbm,bbold}
\usepackage{mathtools}
\usepackage{standalone}
\usepackage{bm}
\usepackage{tensor} 
\usepackage{jheppub}
\usepackage[makeroom]{cancel}
\usepackage{xcolor}
\usepackage{enumerate}
\usepackage{hyperref}

\newcommand{\bfp}{\textbf{p}}

\newcommand{\openone}{\mathbb{1}}

\abstract{In quantum field theory, particle creation occurs, in general, when an intense external field, such as an electromagnetic field, breaks time translational invariance. This leads to an ambiguity in the definition of the vacuum state. In cosmological backgrounds this ambiguity has been reduced by imposing that the quantization preserves the symmetries of the system and that the dynamics is unitarily implemented. In this work, we apply these requirements to the quantization of a massive charged fermionic field coupled to a classical time-dependent homogeneous electric field, extending previous studies done for a scalar field. We characterize the quantizations fulfilling the criteria above and we show that they form a unique equivalence class of unitarily related quantizations, which provide a well-defined number of created particles at all finite times.}
 
\begin{document} 
\maketitle
\flushbottom

\section{Introduction}

Quantum Electrodynamics is a theory experimentally verified with a high level of accuracy, mainly through the study of perturbative effects observed for example in particle-particle collisions. Nonetheless, Quantum Field Theory (QFT) allows to make predictions beyond this perturbative regime. This is the case, for instance, for the creation of particle-antiparticle pairs from matter fields coupled to intense electromagnetic fields: the so-called Schwinger effect. This phenomenon was first suggested by F. Sauter \cite{Sauter1931742}, although it carries the name of Schwinger as he was the one who first explained it in the context of Quantum Electrodynamics for slowly varying fields \cite{Schwinger1951}.

This kind of phenomena of creation of particles is, in general, typical of situations in which the vacuum state of a quantum field is strongly affected by an external agent. In QFT in curved spacetimes the curvature of the spacetime plays the role of this agent, as it happens, for example, in the Hawking effect, which describes the radiation of a black hole \cite{BlackHexplosons}. The problem is that the direct experimental verification of these tiny effects of creation of particles caused by the gravitational field are still unreachable (although quantum acoustic Hawking radiation from analogue black holes in atomic Bose-Einstein condensates has already been observed \cite{Steinhauer_2016}). Analogously, in order to verify empirically the Schwinger effect it is necessary to generate electromagnetic fields above the Schwinger limit, which is of the order of $10^{18} \ \mathrm{V/m}$ \cite{Yakimenko_2019}, with all the technical difficulties that dealing with such intense fields implies. However, there are experimental proposals, such as the ones based on ultraintense lasers \cite{Lasers}, which may make the Schwinger effect one of the first non-perturbative phenomena tested. Thus, it is a unique opportunity for studying the non-perturbative regime in QFT.

In the process of canonical quantization of a free field in Minkowski spacetime, it is usual to unitarily implement the invariance of the classical system under the Poincaré symmetry in the quantum theory. This selects a unique set of annihilation and creation operators. Thus, there is no ambiguity in the definition of the quantum vacuum and unique notions of particles and antiparticles exist.

Nevertheless, we should keep in mind that QFT is a theory in terms of fields and not of particles. Indeed, in general curved spacetimes this particle interpretation might not even exist. The key is that for general geometries the classical system may not be invariant under so many symmetries or none at all. In particular, particle creation effects occur in general when the external agent breaks the invariance under time translations, so that there is not a preferred vacuum, but infinite different choices which, in fact, could give rise to non-equivalent quantum theories. 

While the motivation for this work comes from the study of QFT in curved spacetimes, in the context of the Schwinger effect we consider a flat background. In this case, an intense electromagnetic field coupled to the matter fields is responsible for breaking the Poincaré symmetry of Minkowski spacetime, implying the creation of particles throughout the evolution of the vacuum.

With the aim of reducing the ambiguity in the choice of vacuum, it is desirable to find other physically reasonable requirements to impose on the quantizations. In the context of fields propagating through homogeneous cosmologies a proposal has been put forward which allows to select a unique family of unitarily equivalent Fock representations and, therefore, physically equivalent quantizations \cite{Cortez:2019orm,Cortez:2020rla}: the unitary implementation of both the classical symmetries of the equations of motion and the quantum field dynamics at all times.

Invariance of the classical system under any symmetry transformation implies in the quantum theory that, if the vacuum remains invariant, it is possible to implement the symmetry transformation unitarily. When such transformations are not symmetries or the vacuum is not invariant, one can still try to impose a weaker condition, namely a unitary (non-trivial) implementation of such transformations. This, in particular, applies to time translations. For a deeper discussion on this point, we refer the reader to \cite{universe7080299}. If the unitarity of the dynamics is imposed, the quantizations at each time, which might not be a priori equivalent, are assured to provide the same physics. This requirement in QFT reminds us of what we do in Quantum Mechanics when we work equally in Schrödinger, interaction and Heisenberg pictures. All of them are known to be related by a unitary evolution operator, which allows us to go from one picture to another rendering equivalent physical descriptions. In addition, for particle creation phenomena such as the Schwinger effect, the requirement of unitarily implementing the dynamics is equivalent to having a well-defined number of generated particles at all finite times.

Unitary dynamics at all finite times is a stronger condition than the usual one found in the literature \cite{Gavrilov:1996pz,WALD1979490}, which only imposes that the states in the asymptotic past and future (when the external agent stops being coupled to the matter fields) are connected unitarily by the $S$-matrix. Actually, these approaches do not succeed in reducing the ambiguity in the definition of the number of particles created in the process during the interaction between matter and external fields. Moreover, in spite of the existence of the $S$-matrix for general backgrounds \cite{WALD1979490}, there is no certainty that there exist asymptotic free-particle states in non-trivial backgrounds \cite{wald1994quantum}. 

The aim of this work is to study the canonical quantization of a massive charged Dirac field in a flat spacetime coupled to a homogeneous time-dependent electric field, extending to the fermionic case the results achieved in \cite{Garay2020} for the scalar field. In additon, we will consider the electric field to be intense enough so that the test matter field does not modify its dynamics; i.e., we will neglect the backreaction. Thus, the electric field will be classical, external and non-dynamical. The present system has already been studied following diverse approaches reviewed in \cite{Gelis_2016}, such as the quantum kinetic equation \cite{Smolyansky1997DynamicalDO,Fedotov_2011}, or the Wigner formalism \cite{Hebenstreit:2010cc,Schwingerwigner}. Here we follow a different procedure.

The main result of our work is the characterization of the possible Fock representations which unitarily implement both the symmetries of the system and the dynamics. Furthermore, we prove that the ambiguity in the quantization is reduced so that they are all unitarily equivalent and, therefore, lead to the same transition amplitudes. In particular, we identify the quantizations with a well-defined total number of particles and antiparticles created throughout the evolution of the vacuum at any instant of time. Nevertheless, the particular definition for this number of particles will still depend on the specific vacuum chosen within the unitary equivalence class.

On the other hand, we will also show that, in order for a quantization to be part of this unitary equivalence class, it must explicitly depend on the electromagnetic potential externally applied to the fermionic field. In particular, this implies that in this non-trivial background the usual Minkowski quantization (fixed by the Poincaré symmetry) does not unitarily implement the dynamics. This is, in fact, in agreement with previous works \cite{Ruijsencharged}.

In section \ref{sec_theoreticalframework}, we describe the canonical quantization of a charged fermionic field in a general electromagnetic background, highlighting the ambiguity in the split of the Hilbert space in one-particle and one-antiparticle sectors. In section \ref{sec_bogoliubov}, we present the concepts of unitary implementation of time-dependent canonical transformations between different Fock representations, also known as Bogoliubov transformations. In section \ref{section_fermionic}, we start particularizing our previous results to the case of a charged fermionic field coupled to a homogeneous electric field. Particularly, we propose that gauge transformations be trivially implemented to keep homogeneity. In section \ref{sec_uniquenessquantization} we prove that the unitary implementation of the dynamics, together with the symmetries of the system, reduces the ambiguity in the quantization to a unique equivalence class. Finally, in section \ref{sec_conclusions} we present a summary and a discussion of the main results of this work.

\section{Fermionic canonical quantization} \label{sec_theoreticalframework}

Our system consists of a fermionic field described by a Dirac spinor $\psi$ of mass $m$ and non-zero charge $q$ coupled to an electromagnetic field in Minkowski spacetime. The starting point of our analysis is its action, namely
\begin{equation}
   S=\int d^4x \ \Bar{\psi}\left[ i\gamma^\mu(\partial_\mu+iqA_\mu)-m\right] \psi,
   \label{fermionicaction}
\end{equation}
where $x=(t,\textbf{x})$, $A_{\mu}$ is the four-vector potential of the external electromagnetic field (which will remain classical all along), $\gamma^\mu$ are the Dirac matrices, and the bar denotes Dirac adjoint ($\Bar{\psi}=\psi^\dagger\gamma^0$).

Spinors are half integral representations of the Lorentz group. In order to capture the fermionic nature of this field, we will consider the components of $\psi$ and $\bar{\psi}$ to be Grassmann anticommuting variables. Consequently, in order to calculate the Euler-Lagrange equations, the derivatives of the action $S$ should be either left or right Grassmann derivatives. We will choose left Grassmann derivatives, which are defined as derivatives of an expression where the variable with respect to which the expression is being differentiated occupies the first place in multiplicative order. Thus, the Dirac equation of motion is
\begin{equation}
    \left[i\gamma^\mu(\partial_\mu+iqA_\mu)-m\right]\psi=0.
    \label{eqdirac}
\end{equation}

One way to proceed in the description of the classical theory is the canonical approach. The canonical phase space of the system is the set of pairs composed by a field and its conjugate momentum $(\psi_0(\textbf{x}),\pi_0(\textbf{x}))$, defined on a Cauchy surface $\Sigma_{t_0}$. Each pair of functions is assumed to be smooth and represents one possible set of initial conditions for the solutions to the equations of motion, $(\psi(x),\pi(x)$). By definition, this canonical phase space is also endowed with a symmetric Poisson bracket structure. For two Grassmann variables $A$ and $B$ it is defined as
\begin{align}
    \poissonbracket{A}{B}_{\text{G}}=-\sum_{\alpha}\left(\partial^{\text{G}}_{\theta^\alpha}A\partial^{\text{G}}_{\pi_\alpha}B+\partial^{\text{G}}_{\theta^\alpha}B\partial^{\text{G}}_{\pi_\alpha}A\right),
    \label{poissonbracket}
\end{align}
where $\partial^{\text{G}}_{\theta^{\alpha}}$ and $\partial^{\text{G}}_{\pi_{\alpha}}$ denote the left Grassmann derivatives with respect to the canonical Grassmann variable $\theta^{\alpha}$ and its conjugate momentum $\pi_{\alpha}$, respectively \cite{Casalbuoni1976}.

The Dirac equation \eqref{eqdirac} has a well-posed initial value formulation, in the sense that for every pair $(\psi_0,\pi_{0})$ there exists a unique smooth solution $\psi$ on the whole spacetime manifold satisfying the initial conditions $\psi|_{\Sigma_{t_0}}=\psi_0$ and $\pi|_{\Sigma_{t_0}}=-i\psi^{\dagger}|_{\Sigma_{t_0}} = \pi_{0}$. Therefore, we can identify this canonical phase space with the so-called covariant phase space: the vector space $\mathcal{S}$ of classical solutions of the Dirac equation with smooth initial data. 

Using integration by parts it can be shown that the form on $\mathcal{S}$ given by 
\begin{align}
Q(\psi_1,\psi_2)=\int_{\Sigma_t} d^3\textbf{x} \  \psi_1^{\dagger}(x)\psi_2(x)
\end{align}
does not depend on the Cauchy surface $\Sigma_t$ on which it is evaluated, i.e., it is time independent. Moreover, $Q$ is a positive-definite inner product, which endows the space of solutions $\mathcal{S}$ with a complex Hilbert space structure. We find here the fundamental difference with respect to the case of a charged scalar field in an electromagnetic background: the space of solutions of the Klein-Gordon field coupled to an external electromagnetic potential has no natural inner product defined on it, but an antisymmetric symplectic form  which fails to be positive definite \cite{wald1994quantum,Garay2020}. In that case, the introduction of a complex structure is necessary in order to construct a Hilbert space of solutions. 

\subsection{One-particle and one-antiparticle Hilbert spaces} \label{sec_split}

Despite the fact that in the fermionic case $\mathcal{S}$ is directly a Hilbert space of solutions, in order to quantize the theory we will define a separable Hilbert space of particles and antiparticles, for which the introduction of a complex structure is necessary.

Indeed, the freedom in defining the one-particle Hilbert space is completely characterized by a choice of complex structure $J:\mathcal{S}\rightarrow\mathcal{S}$, which by definition is a real antihermitian operator satisfying $J^2=-\openone$. It defines orthogonal projection operators
\begin{equation}
P^{\pm}=(\openone\mp iJ)/2
\end{equation}
on the two spectral eigenspaces of $J$ with eigenvalues $\pm i$.

We define the particle states as the ones associated with the eigenvalue $+i$. Thereby, the states associated with $-i$ could be viewed as \textit{holes} which represent the antiparticle states. In this way, the one-particle Hilbert space $\mathcal{H}^{\text{p}}$ can then be identified with the Cauchy completion of $P^+\mathcal{S}$ with respect to $Q$. The choice of $\mathcal{H}^{\text{p}}$ determines unequivocally the  one-antiparticle Hilbert space, $\mathcal{H}^{\text{ap}}$, whose elements are called antiparticle states. If $\psi\in P^{+}\mathcal{S}$ is a particle, $\psi^*\in P^{-}\mathcal{S}$ is its antiparticle, but viewed as a hole. In order to see $\psi^*$ as an antiparticle state, $\mathcal{H}^{\text{ap}}$ is taken as the subspace of $\mathcal{S}^*$ associated with the eigenvalue $+i$ of $J$. In other words, $\mathcal{H}^{\text{ap}}$ is defined as the Cauchy completion (with respect to $Q^*$) of $P^+\mathcal{S}^*$. In fact, as a matter of consistency, it is easy to see that $P^+\mathcal{S}^*=(P^-\mathcal{S})^*$. The complete Hilbert space is then $\mathcal{H}=\mathcal{H}^{\text{p}}\oplus \mathcal{H}^{\text{ap}}$. Let us remark here that $\mathcal{H}$ is not the Hilbert space of solutions $\mathcal{S}$. In fact, $\mathcal{S}=\mathcal{H}^{\text{p}}\oplus (\mathcal{H}^{\text{ap}})^*$, whereas $\mathcal{H}=\mathcal{H}^{\text{p}}\oplus \mathcal{H}^{\text{ap}} \subset \mathcal{S}\oplus \mathcal{S}^*$.

We can now choose orthonormal bases $\{\psi_n^{\text{p}}\}\subset \mathcal{H}^{\text{p}}$, $\{(\psi_n^{\text{ap}})^*\}\subset \mathcal{H}^{\text{ap}}$, so that $\{\psi_n^{\text{p}},(\psi_n^{\text{ap}})^*\}$ is an orthonormal basis of $\mathcal{H}$. Consequently, for every solution $\Psi \in \mathcal{S}$ there exist unique complex coefficients $a_n$ and $b_n^*$ such that
\begin{equation}
    \Psi(x)=\sum_n \big[ a_n\psi_n^{\text{p}}(x)+b_n^*\psi_n^{\text{ap}}(x) \big].
    \label{Psipap}
\end{equation} 
These coefficients, which are associated with the complex structure $J$, are called annihilation and creation variables, respectively. Here we consider the label $n$ to be discrete. In the case that $n$ were a continuous index, all equations would be naturally written with integrals instead of summations.

It is easy to see that the symmetric Poisson bracket structure \eqref{poissonbracket} induces the following algebra for the creation and annihilation variables:
\begin{align}
    \{a_n,a_m^*\}_{\text{G}}=&\{b_n,b_m^*\}_{\text{G}}=-i\delta_{n,m},
    \label{poissoncndn}
\end{align}
the rest of Poisson brackets among them being zero.

\subsection{Canonical quantization}

In order to define the quantum theory, the full Hilbert space is chosen to be the antisymmetric Fock space,
\begin{equation}
    \mathcal{F}=\oplus_{n=0}^{\infty}\left( \otimes^n_{\text{A}} \mathcal{H} \right),
\end{equation}
where $\otimes^n_{\text{A}} \mathcal{H}$ is the antisymmetric tensor product of $n$ copies of $\mathcal{H}$. The complex coefficients $a_n$ and $b_n$  are mapped to annihilation operators $\hat{a}_n,\hat{b}_n$ acting on the Fock space, verifying the canonical anticommutation relations obtained by the prescription 
\begin{equation}
    \{\cdot,\cdot\}_{\text{G}}\rightarrow \{\hat{\cdot},\hat{\cdot}\}=i\widehat{\{{\cdot},{\cdot}\}}_{\text{G}},
\end{equation}
where $\{\hat{\cdot},\hat{\cdot}\}$ is the anticommutator. The only non-vanishing relations among them are therefore
\begin{equation} 
    \{\hat{a}_n,\hat{a}_m^{\dagger}\}=\{\hat{b}_n,\hat{b}_m^{\dagger}\}=\delta_{n,m}.
\end{equation}
Furthermore, we define the Fock vacuum as the state which is annihilated by all annihilation operators $\hat{a}_n$ and $\hat{b}_n$. Finally, the quantum field operator $\hat{\Psi}$ on the Fock space is defined in the Heisenberg picture by simply substituting the coefficients $a_n$ and $b_n$ by their corresponding operators in \eqref{Psipap}; i.e.,
\begin{equation}
    \hat{\Psi}(x)=\sum_n \big[ \hat{a}_n\psi_n^{\text{p}}(x)+\hat{b}_n^{\dagger}\psi_n^{\text{ap}}(x) \big].
    \label{Psipapquantized}
\end{equation} 

In summary, every complex structure $J$ defines a split of the space of solutions $\mathcal S$ which leads to two one-particle Hilbert spaces, namely $\mathcal{H}^{\text{p}}$ and $\mathcal{H}^{\text{ap}}$. Therefore, the definitions of what we call particles (elements of $\mathcal{H}^{\text{p}}$) and antiparticles (elements of $\mathcal{H}^{\text{ap}}$) are characterized by the complex structure, which keeps all the information of the particular quantization chosen for the classical theory.

\section{Bogoliubov transformations} \label{sec_bogoliubov}

In the definition of the one-particle Hilbert space $\mathcal{H}^{\text{p}}$ and, therefore, in the selection of the annihilation and creation operators of the quantum theory, there is an ambiguity based on the choice of the complex structure $J$. We are interested in finding whether some complex structures are more adequate than others in order to quantize our theory. 

One requirement which is reasonable to impose is that $J$ defines a quantization which unitarily implements the symmetries of the classical system. This condition is extremely restrictive in certain cases, for instance for free quantum fields in Minkowski spacetime: if the Poincaré symmetry is unitarily implemented in the quantum theory, the ambiguity in the selection of the complex structure disappears and a unique complex structure is selected. Thereby, there is a preferred vacuum and well-defined notions of particle-antiparticle can be provided. However, for general geometries, or even for Minkowski spacetime with a non-trivial background (as an electromagnetic potential coupled to the fields), the classical theory is not invariant under such group of transformations, and consequently, imposing symmetries in the quantum theory can reduce the choice of the complex structure but, in general, does not suffice to fix it unequivocally. Then, the concepts of particle and antiparticle inherit the residual ambiguity.

\subsection{Classical Bogoliubov transformations} \label{sec_classicalbog}

With the aim of reducing the ambiguity in the choice of the complex structure, we need to compare different representations of the canonical anticommutation relations and understand whether they can be physically related in some sense under certain conditions. In order to mathematically formulate this problem, we are going to consider canonical transformations of the fields. 
Let us consider a vector space of classical solutions $\mathcal{S}$ and two complex structures on it, $J$ and $\tilde{J}$, determining Hilbert spaces $\mathcal{H}$ and $\tilde{\mathcal{H}}$, respectively, with the corresponding orthonormal bases $\{\psi^{\text{p}}_n,(\psi^{\text{ap}}_n)^*\} \subset \mathcal{H}$ and $\{\tilde{\psi}^{\text{p}}_n,(\tilde{\psi}^{\text{ap}}_n)^*\}\subset \tilde{\mathcal{H}}$. We can write any classical solution $\Psi\in\mathcal{S}$ in terms of both bases: 
\begin{equation}
    \Psi=\sum_n \big( a_n\psi_n^{\text{p}}+b_n^*\psi_n^{\text{ap}} \big)
    =\sum_n \big( \tilde{a}_n\tilde{\psi}_n^{\text{p}}+\tilde{b}_n^*\tilde{\psi}_n^{\text{ap}}\big).
    \label{Psicctilde}
\end{equation} 
Classically, $\Psi$ is the same solution expressed in different bases. Nevertheless, since what we quantize are the annihilation and creation variables $\{a_n,b^*_n\}$, $\{\tilde{a}_n,\tilde{b}^*_n\}$, which are different for $J$ and $\tilde{J}$, both representations will lead to different field operators.

A canonical transformation can be written as
\begin{equation}
    \mqty(\tilde{a}_n\\\tilde{b}^*_n)=\sum_m\mqty(\alpha_{nm}^{\text{p}} & \beta_{nm}^{\text{p}} \\ \beta_{nm}^{\text{ap}} & \alpha_{nm}^{\text{ap}})\mqty(a_m \\ b_m^*),
    \label{cndnbogoliubov}
\end{equation}
with the matrix entries satisfying appropriate relations so that the Poisson algebra of the creation and annihilation variables is preserved. Any canonical transformation of this sort is called a classical Bogoliubov transformation \cite{wald1994quantum}.

The concepts of particle and antiparticle are in general different for both complex structures: the coefficients $\beta_{nm}^{\text{p}}$ mix the states in $\tilde{\mathcal{H}}^{\text{p}}$ with the states in $\mathcal{H}^{\text{ap}}$; and $\beta^{\text{ap}}_{nm}$ mix $\tilde{\mathcal{H}}^{\text{ap}}$ with $\mathcal{H}^{\text{p}}$. Only if these $\beta$-coefficients vanished the definitions of particle and antiparticle would be the same for $J$ and $\tilde{J}$. In fact, this trivial transformation would correspond to independent changes of bases in the one-particle and in the one-antiparticle Hilbert spaces, not modifying the quantization of the classical theory.

Note that, in general, a classical time-dependent Bogoliubov transformation $B$ can connect two isomorphic vector spaces of solutions $\mathcal{S}$ and $\mathcal{S}'$. For example, as we are going to see in section \ref{sec_gauge}, the Dirac equation is not invariant under a $U(1)$ gauge transformation, but covariant. Nevertheless, when choosing a complex structure $\tilde{J}'$ on $\mathcal{S}'$, it is always possible to write the corresponding complex structure on $\mathcal{S}$: 
\begin{equation}
    \tilde{J}=B^{-1}\tilde{J}'B.
\end{equation}
Then, without loss of generality we will assume in the following that every complex structure is defined on the domain of the Bogoliubov transformation.

\subsection{Quantum Bogoliubov transformations} \label{sec_unitary}

Let $B:\mathcal{S}\rightarrow\mathcal{S}'$ be a classical (time-dependent) Bogoliubov transformation. Two complex structures $J$ and $\tilde{J}=B^{-1}\tilde{J}'B$ on $\mathcal{S}$ provide quantum field operators $\hat{\Psi}$ and $\tilde{\hat{\Psi}}$, respectively, associated with a classical solution $\Psi\in\mathcal{S}$. $B$ is said to be unitarily implementable in the quantum theory if and only if it can be represented by a unitary operator $\hat{U}_B:\mathcal{F} \rightarrow \tilde{\mathcal{F}}$ such that 
\begin{equation}
    \tilde{\hat{\Psi}}=\hat{U}_B\hat{\Psi}\hat{U}_B^{-1}.
\end{equation}
In this case, $J$ and $\tilde{J}$ are said to be unitarily equivalent, and the states $\ket{\phi}\in \mathcal{F}$ and $|\tilde{\phi}\rangle=\hat{U}_B\ket{\phi}\in \tilde{\mathcal{F}}$ lead to the same transition amplitudes; i.e., $\langle\tilde{\phi}_1|\tilde{\hat{\Psi}}(x)|\tilde{\phi}_2\rangle=\bra{\phi_1}\hat{\Psi}(x)\ket{\phi_2}$. 

A rather useful characterization of the condition of unitary equivalence between two complex structures was given by Shale \cite{Shale:1962,RUIJSENAARS1978105}: the necessary and sufficient condition for a classical Bogoliubov transformation to be unitarily implementable in the quantum theory is that its $\beta$-Bogoliubov coefficients are Hilbert-Schmidt; namely, that they satisfy
\begin{equation}
    \sum_{n,m} \left( |\beta_{nm}^p|^2+|\beta_{nm}^{ap}|^2 \right) < \infty.
    \label{betaconvergent}
\end{equation}
As we are considering general time-dependent Bogoliubov transformations, this condition has to be satisfied at all finite times.

For systems with finite number of degrees of freedom, the sum in \eqref{betaconvergent} is trivially finite, which proves that all quantizations are unitarily equivalent, in agreement with the Stone-von Neumann theorem \cite{wald1994quantum}. However, in the infinite dimensional case, as we are going to see later, there exist Bogoliubov transformations which cannot be implemented as unitary operators, so there exist unitarily inequivalent quantizations. Consequently, the selection of the complex structure plays a relevant role in the process of quantization, and some effort should be made to distinguish which ones are more appropriate in each case.

Thus, we need to impose physical criteria on the complex structure in order to reduce the ambiguity in the quantization. Motivated by previous studies in cosmology \cite{Cortez:2019orm,Cortez:2020rla} and in the Schwinger effect for a charged scalar field \cite{Garay2020}, our central work in section \ref{sec_uniquenessquantization} will be to characterize the complex structures which preserve the symmetries of the system and unitarily implement the dynamical evolution of a charged fermionic field in presence of a homogeneous time-dependent electromagnetic background. Therefore, let us describe how can time evolution be treated as a Bogoliubov transformation.

\subsection{Time evolution as a Bogoliubov transformation} \label{timeevolution}

Let us review here some results which will be useful in our later discussion. For more details, see \cite{Cortez:2019orm,Cortez:2020rla,CORTEZ201536}.

Because of the well-posed initial value formulation of the Dirac equation, time evolution from time $t_0$ to time $t$ of a classical solution $\Psi\in\mathcal{S}$ is described by a time-dependent classical Bogoliubov transformation $T(t_0,t):\mathcal{S}\to \mathcal{S}$. With this transformation we are allowing the annihilation and creation operators to be time-dependent, distributing the dynamics between them and the elements of the considered orthonormal bases in the expansion \eqref{Psipapquantized}. This is analogous to working in different pictures in Quantum Mechanics.

Given a complex structure $J_{t_0}$, $T(t_0,t)$ defines the one-parameter family of complex structures
\begin{equation}
    J_t=T(t_0,t)J_{t_0}T^{-1}(t_0,t).
\end{equation}
If we impose that the Bogoliubov transformation $T(t_0,t)$ is unitarily implementable for all $t$, then $J_t$ will be unitarily equivalent to $J_{t_0}$ for all $t$. In other words, quantizations will be unitarily equivalent, and, consequently, will provide the same physics during the evolution of the fields. This unitary implementation of the dynamics is exactly what we are going to demand to the quantization of a charged fermionic field in an electromagnetic background in section \ref{sec_unitarydynamics}. Later, in section \ref{sec_uniqueness} we will prove that this reduces the ambiguity in the selection of complex structures to a unique class that defines unitarily equivalent quantizations.

\subsection{Gauge transformations} \label{sec_gauge}

Let us consider local $U(1)$ gauge transformations. They define a time-dependent classical Bogoliubov transformation $G(g):\mathcal{S}_A \rightarrow \mathcal{S}_{A^g}$ given by 
\begin{equation}
    \Psi \mapsto G(g)\Psi=e^{iqg(x)}\Psi,
\end{equation}
where $\mathcal{S}_A$ is the vector space of classical solutions to the Dirac equation \eqref{eqdirac} with the four-vector potential $A_{\mu}$, and $g$ is a general function. The Dirac equation is gauge covariant but not invariant since $A_{\mu}$ transforms as $
    A_{\mu} \rightarrow A_{\mu}^g=A_{\mu}-\partial_{\mu}g$. 
This is due to the fact that $A_{\mu}$ is a nondynamical gauge field, externally imposed on the equations of motion for the matter fields $\psi$. Therefore, as the Dirac equation depends on the particular choice of the gauge field $A_{\mu}$, so does its space of solutions, meaning that, in general, $\mathcal{S}_A$ and $\mathcal{S}_{A^g}$ do not coincide.

Given a complex structure $J$ on $\mathcal{S}_A$, a gauge transformation  $G(g)$ defines another complex structure 
\begin{equation}
    J_g=G(g)JG(g)^{-1}
\end{equation}
on $\mathcal{S}_{A^g}$
that trivially implements $G(g)$.
Indeed, gauge transformations simply act by multiplication by a phase. This translates into a diagonal Bogoliubov matrix with null $\beta$-coefficients. Thus, gauge transformations can always be unitarily implemented in the quantum theory by an operator which acts on the quantum fields simply by multiplication of~$e^{iqg(x)}$. In particular, as we are going to see in section \ref{sec_symmetries}, the homogeneity that we will impose on the electric field applied to the Dirac field will select a privileged gauge (the temporal gauge) which preserves the homogeneity of the Dirac equation. We will work in this fixed gauge. To translate the results obtained to other gauges, we simply need to classically transform the fields.

\section{Homogeneous  electric field} \label{section_fermionic}

Taking into account all the previous results regarding the canonical quantization of a charged fermionic field in Minkowski spacetime coupled to a general electromagnetic field, let us particularize our study to the case in which this external field is a homogeneous electric field, so no magnetic component is present. 

We would like to benefit from this property, so we will choose a gauge explicitly exhibiting these symmetries in the action. This is the case of the temporal gauge, in which the four-potential takes the form $A_\mu(t,\textbf{x})=(0,\textbf{A}(t))$. Without loss of generality, we also choose the potential to be parallel to the $z$-axis, i.e., the third spatial direction: $\textbf{A}(t)=(0,0,A(t))$. Finally, to completely fix the gauge we are going to set $A(t_0)=0$.

\subsection{Mode decomposition}

Since the temporal gauge makes the equation of motion invariant under spatial translations, we can expand its solution in plane wave modes $\psi_{\textbf{p}}(t)$, one for each wave vector $\textbf{p}\in \mathbb{R}^3$:
\begin{equation}
    \psi(t,\textbf{x})=\int_{\mathbb{R}^3} \frac{d^3\textbf{p}}{(2\pi)^{3/2}} \ e^{i\textbf{p}\cdot \textbf{x}}\psi_\textbf{p}(t).
    \label{modedecomposition}
\end{equation}
These modes have decoupled actions, 
\begin{equation}
    S=\int_{\mathbb{R}^3} d^3\bfp \ S_{\bfp},
\end{equation}
with
\begin{equation} 
    S_{\bfp}=\int_{\mathbb{R}} dt \  \Bar{\psi}_{\bfp}(t)\gamma^0 \Big[ i\partial_t -\left(p_1\gamma^0\gamma^1+p_2\gamma^0\gamma^2+m\gamma^0\right)-\left(p_3+qA(t)\right)\gamma^0\gamma^3 \Big] \psi_{\textbf{p}}(t).
    \label{actionmodedecomposition}
\end{equation}
It is now convenient to decompose the solution in eigenvectors of $\gamma^0\gamma^3$. The eigenvalues of $\gamma^0\gamma^3$ are +1 and -1, each of them with double degeneracy. Two eigenvectors which form an orthonormal basis of the subspace associated with the eigenvalue $+1$ (in the Dirac representation of the Dirac matrices) are
\begin{equation}
    R_1= \frac{1}{\sqrt{2}}(1,0,1,0)^\textsc{t}, \quad R_2=\frac{1}{\sqrt{2}}(0,-1,0,1)^\textsc{t},
\end{equation}
where $\textsc{t}$ denotes transposition. Besides, the vectors
\begin{equation}
   -\omega_{\perp}^{-1}\left(p_1\gamma^0\gamma^1+p_2\gamma^0\gamma^2-m\gamma^0\right)R_r,
   \label{basisR}
\end{equation}
with $r=1,2$, form an orthonormal basis on the subspace with eigenvalue $-1$, where
\begin{equation} \label{omegaperp}
\omega_{\perp}=\sqrt{p_1^2+p_2^2+m^2}.
\end{equation} 
Thus, we can write each mode $\psi_{\textbf{p}}$ as a linear combination of these four vectors,
\begin{equation}
    \psi_{\textbf{p}}(t)=\gamma^0\sum_{r=1,2}\big[\sigma^*_{r,\textbf{p}}(t) -\omega_{\perp}^{-1}\left(p_1\gamma^0\gamma^1+p_2\gamma^0\gamma^2-m\gamma^0\right)\chi_{r,\textbf{p}}(t)\big]R_r,
    \label{eachmodedecomposition}
\end{equation}
for some time-dependent scalar fields $\sigma^*_{r,\textbf{p}}(t)$, $\chi_{r,\textbf{p}}(t)$. Since $\psi$ and $\Bar{\psi}$ are Grassmann variables, their anti-commuting nature is now inherited by the scalar functions $\sigma_{r,\textbf{p}}$ and $\chi_{r,\textbf{p}}$. The prefactor $\gamma^0$ has been introduced for convenience.

Using this form for each mode in expression \eqref{modedecomposition} and inserting everything in \eqref{actionmodedecomposition}, the action of the system becomes
\begin{equation}
    S=\sum_{r=1,2}\int_{\mathbb{R}^3}d^3\textbf{p} \ S_{r,\bfp},
\end{equation}
with
\begin{equation} \label{actionsigmachi}
\begin{split}
    S_{r,\bfp}&=\int_{\mathbb{R}} dt  \  \big[ i\left(\chi_{r,\textbf{p}}^*\dot{\chi}_{r,\textbf{p}}+\sigma_{r,\textbf{p}}\dot{\sigma}_{r,\textbf{p}}^*\right)
    +(p_3+qA)\left(\sigma_{r,\textbf{p}}\sigma_{r,\textbf{p}}^*-\chi_{r,\textbf{p}}^*\chi_{r,\textbf{p}}\right)
    \\
    &-\omega_{\perp}\left(\chi_{r,\textbf{p}}^*\sigma_{r,\textbf{p}}^*+\sigma_{r,\textbf{p}}\chi_{r,\textbf{p}}\right)\big].
    \end{split}
\end{equation}
The only non-vanishing symmetric Poisson brackets among the canonical variables are:
\begin{equation}
    \poissonbracket{\sigma_{r,\textbf{p}}}{\sigma_{s,\textbf{q}}^*}_{\text{G}}=\poissonbracket{\chi_{r,\bfp}}{\chi^*_{s,\textbf{q}}}_{\text{G}}=-i\delta_{r,s}\delta^{(3)}(\bfp-\textbf{q}).
    \label{poissonbrackets}
\end{equation} 
In addition, we see from \eqref{actionsigmachi} that the modes $\sigma^*_{r,\textbf{p}}(t)$ have the same dynamics for $r=1$ and $r=2$, which allows us to drop the index $r$. Furthermore, the modes labelled by $\textbf{p}$ are decoupled from one another, which allows us to drop the index $\textbf{p}$ (keeping in mind that the equations of motion depend on the momentum). This also applies to $\chi_{r,\textbf{p}}(t)$. Using left Grassmann variational derivatives we obtain that both $\chi$ and $\sigma$ satisfy harmonic oscillator equations with time-dependent complex frequency $\omega(t)^2+iq\dot{A}(t)$, where
\begin{equation}
    \omega(t)=\sqrt{\omega_{\perp}^2+[p_3+qA(t)]^2}.
    \label{omega}
\end{equation}
Furthermore, $\chi$ determines the value of $\sigma^*$. The equations of motion can then be written as
\begin{align}
    &\ddot{\chi}+\left(\omega^2+iq\dot{A}\right)\chi=0,
    \label{soeqchi} \\
    & \sigma^*=\omega_{\perp}^{-1}\left[ i\dot{\chi}-(p_3+qA)\chi \right]. \label{sigma}
\end{align}
Note that at $t=t_0$, due to the condition $A(t_0)=0$ we get that 
\begin{equation}
    \omega(0)=\sqrt{m^2+p^2}=\omega_0,
\end{equation}
where $p=\abs{\mathbf{p}}$.

\subsection{Classical time evolution}

We are interested in obtaining the time evolution of the coupled modes $\chi(t)\equiv \chi_{r,\textbf{p}}(t)$ and $\sigma^*(t)\equiv\sigma^*_{r,\textbf{p}}(t)$ in terms of the initial conditions $\chi(t_0)$ and $\sigma^*(t_0)$. According to section \ref{timeevolution}, this evolution is given by a classical Bogoliubov transformation:
\begin{equation}
    \mqty(\chi(t)\\ \sigma^*(t))={T}(t_0,t)\mqty(\chi(t_0)\\ \sigma^*(t_0)).
    \label{evolutionmodeschi}
\end{equation}

On the other hand, equation \eqref{soeqchi} is a linear second order differential equation with complex coefficients, which means that we can choose a basis of solutions to be
$e^{i(-1)^l\Theta^l(t)}$, with $\Theta^l$ ($l=1,2$) complex functions to be determined. Then, any solution to this equation can be expressed in terms of them as
\begin{equation}
    \chi(t)=C^1 e^{-i\Theta^1(t)}+C^2 e^{i\Theta^2(t)},
    \label{chienfunciondetheta}
\end{equation}
where $C^l\in\mathbb{C}$ are uniquely determined by the initial conditions $\Theta^l(t_0)$ and $\dot{\Theta}^l(t_0)$. By inserting \eqref{chienfunciondetheta} in \eqref{evolutionmodeschi} and using \eqref{sigma} we can deduce the expressions for the entries of the matrix $T(t_0,t)$:
\begin{equation}
\begin{split}
    T^{11}(t_0,t)&=W(t_0)\big[\Omega^2(t_0)e^{-i\delta^1(t)}+\Omega^1(t_0)e^{i\delta^2(t)}\big], \\
    T^{12}(t_0,t)&=W(t_0) \omega_{\perp} \big[e^{-i\delta^1(t)}-e^{i\delta^2(t)}\big],  \\
    T^{21}(t_0,t)&=W(t_0)\omega_{\perp}^{-1} \big[\Omega^2(t_0)\Omega^1(t)e^{-i\delta^1(t)} -\Omega^1(t_0)\Omega^2(t)e^{i\delta^2(t)}\big], \\
    T^{22}(t_0,t)&=W(t_0) \big[\Omega^1(t)e^{-i\delta^1(t)}+\Omega^2(t)e^{i\delta^2(t)}\big], \label{elementsmatrix}
\end{split} 
\end{equation}
where
\begin{align} 
\Omega^l(t)=&\dot{\Theta}^l(t)+(-1)^l(p_3+qA(t)), \label{Omegal} \\
\delta^l(t)=&\Theta^l(t)-\Theta^l(t_0), \\
W(t_0)=&[\Omega^1(t_0)+\Omega^2(t_0)]^{-1}.
\end{align}  

\subsection{Classical solutions in the ultraviolet regime} \label{sec_asymptotic} 

In the following we study the behaviour of the modes $\chi$ and $\sigma^*$ in the ultraviolet (UV) regime, that is, in the asymptotic limit in which the wave number $p$ is large, since the Hilbert-Schmidt condition does not depend on the lower energy scales for a massive Dirac field. For convenience, we write
\begin{equation}
    \Theta^l(t)=\int_{t_0}^t d\tau \ \left[\omega(\tau)+\Lambda^l(\tau)\right].
    \label{formtheta}
\end{equation}
Here we have chosen $\Theta^l(t_0)=0$ without loss of generality. In addition, taking into account that $A(t_0)=0$, we impose $\dot{\Theta}^l(t_0)=\omega_0$, which implies $\Lambda^l(t_0)=0$.

We assume now that $\Lambda^l$ are complex functions which behave asymptotically as $\order{p^{-1}}$, provided that the time-dependence of the electromagnetic potential is controlled in a sense that will be specified below. With the objective of checking the self-consistency of this fact, we insert the solutions $e^{i(-1)^l\Theta^l(t)}$ in \eqref{soeqchi}. We then obtain a Riccati equation for the functions $\Lambda^l$:
\begin{equation}
    \dot{\Lambda}^l=i(-1)^l\left[(\Lambda^l)^2+2\omega\Lambda^l\right]-\dot{\omega}+(-1)^lq\dot{A}.
    \label{ricatti}
\end{equation}
As $\omega=\order{p}$, $\dot{\omega}=\order{1}$, and as we are assuming $\Lambda^l=\order{p^{-1}}$, the term $i(-1)^l(\Lambda^l)^2$ is negligible at large $p$. The solution in the UV regime to the resulting linear equation, after integrating by parts, is
\begin{align}
    \Lambda^l(t)&=\frac{i}{2}(-1)^l\bigg[-\Gamma^l(t)+e^{2i(-1)^l\theta(t)}\left(\Gamma^l(t_0)+\int^t_{t_0} d\tau \ e^{-2i(-1)^l\theta(\tau)}\dot{\Gamma}^l(\tau)\right)\bigg], 
    \label{Lambda}
\end{align}
where
\begin{equation}
    \theta(t)= \int_{t_0}^t d\tau \ \omega(\tau),\quad \Gamma^l(t)=\frac{1}{\omega(t)}\left[\dot{\omega}(t)-(-1)^lq\dot{A}(t)\right].
\end{equation}

Given that $\Gamma^l(t)=\order{p^{-1}}$, it is easy to check that $    \Lambda^l(t)=\order{p^{-1}}$ in the UV regime, as assumed, as long as $\Gamma^l(t)$ and the integral in \eqref{Lambda} are finite. This is a restriction to the time dependence of the possible electromagnetic potentials for which \eqref{formtheta} is valid. From now on we will assume that we are dealing with potentials satisfying this mild requirement. It can be seen that if $\dot{\Gamma}^l$ has a finite number of sign changes in $[t_0,t]$ and $\Gamma^l(t_0)$ and $\Gamma^l(t)$ are finite, this condition is verified. Indeed, with these hypotheses we have that
\begin{equation}
\begin{split}
    &\left|\int^t_{t_0} d\tau \ e^{-2i(-1)^l\theta(\tau)}\dot{\Gamma}^l(\tau)\right| \leq \int^t_{t_0} d\tau \left| \dot{\Gamma}^l(\tau) \right| \\
    &=\left|\sum_{i=1}^n \int_{t_{i-1}}^{t_i} d\tau \ (-1)^{s_i} \dot{\Gamma}^l(\tau)\right|=\left|(-1)^{s_n}\Gamma^l(t)-(-1)^{s_1}\Gamma^l(t_0)\right|
\end{split}
\end{equation}
is finite, where $[t_0,t]=\cup_{i=1}^n[t_{i-1},t_i]$, $\dot{\Gamma}^l(t_j)=0$ ($j=2,...,n-1$), and $s_i$ denotes the sign of $\dot{\Gamma}^l$ in $[t_{i-1},t_i]$. In particular, potentials which turn on and off asymptotically, remaining finite at all times, are of this type. Then, our formalism applies to a very general family of potentials, including as a particular case those associated with electromagnetic fields localized in time. These are the ones usually found in the literature with analytical solutions of the Dirac equation \cite{Gavrilov:1996pz}.

\section{Reduction of the ambiguity in the quantization} \label{sec_uniquenessquantization}

In section \ref{sec_unitary} we discussed that, in general, there exist unitarily inequivalent quantizations. In order to reduce this ambiguity, we also commented that it is reasonable to restrict our study to complex structures which preserve the symmetries of the system and which unitarily implement the dynamics. This is precisely what we are doing in this section for the case of a massive charged fermionic field in presence of a homogeneous time-dependent electric field.

\subsection{Preservation of the symmetries} \label{sec_symmetries}

In the case under study, the preservation of the symmetries implies two conditions. First, our quantization should be explicitly invariant under spatial translations due to the homogeneity of the electromagnetic field. Second, the complex structure should not mix modes with different $(r,\textbf{p})$, which are decoupled in the equations of motion. Both conditions require the temporal gauge fixing condition that we have used, namely, $A_{\mu}(t)=(0,0,0,A(t))$.

Nevertheless, $\chi_{r,\textbf{p}}$ and $\sigma^*_{r,\textbf{p}}$ remain dynamically coupled by equation \eqref{sigma}. Therefore, the annihilation and creation variables $a\equiv a_{r,\textbf{p}}$ and $b^*\equiv b^*_{r,\textbf{p}}$, which are in general time-dependent, will be given by a linear combination of the modes $\chi\equiv \chi_{r,\textbf{p}}$ and $\sigma^*\equiv\sigma^*_{r,\textbf{p}}$ exclusively: 
\begin{equation}
    \mqty(a(t)\\b^*(t))=\mathfrak{J}(t)   \mqty(\chi(t)\\\sigma^*(t)), \quad \mathfrak{J}(t)=\mqty(f_1(t)&f_2(t)\\g_1(t)&g_2(t)).
\label{abest}
\end{equation}
$\mathfrak{J}(t)$ is a matrix of time-dependent functions parametrizing all the possible complex structures compatible with our requirements. These functions need to verify some conditions between them in order to represent a complex structure. Indeed, the modes $\chi$ and $\sigma^*$ and the creation and annihilation variables $a$ and $b$ must satisfy \eqref{poissonbrackets} and the classical symmetric Poisson bracket relations \eqref{poissoncndn}, rewritten as
\begin{equation}
    \poissonbracket{a_{r,\bfp}}{a^*_{s,\textbf{q}}}_G=\poissonbracket{b_{r,\bfp}}{b^*_{s,\textbf{q}}}_G=-i\delta_{r,s}\delta^{(3)}(\bfp-\textbf{q}).
\end{equation}
This requirement implies the following restrictions:
\begin{gather}
    |f_1(t)|^2+|f_2(t)|^2=1,
        \label{idf1f2.a} \\
    g_1(t)=f^*_2(t)e^{i\kappa(t)}, \quad g_2(t)=-f^*_1(t)e^{i\kappa(t)},
    \label{relationsbea}
\end{gather}
where $\kappa(t)$ is an arbitrary real function. The choice of particular $f_i$ and $g_i$ ($i=1,2$) determines the definition of the creation and annihilation variables via \eqref{abest}, and therefore, selects the quantization prescription unequivocally.

\subsection{Unitary implementation of the dynamics} \label{sec_unitarydynamics}

The second requirement that we are imposing on the physically relevant complex structures is that they unitarily implement the time evolution. This will restrict the choice of $f_i$ and $g_i$.

As said in section \ref{timeevolution}, the evolution from time $t_0$ to time $t$ of the modes $(\chi,\sigma^*)$, for each $(r,\textbf{p})$, is determined by a time-dependent classical Bogoliubov transformation of the type \eqref{cndnbogoliubov}:
\begin{equation}
    \mqty(a(t)\\b^*(t))=\mathfrak{B}(t_0,t)   \mqty(a(t_0)\\b^*(t_0)), \quad\mathfrak{B}(t_0,t)=\mqty(\alpha^f(t_0,t)&\beta^f(t_0,t)\\\beta^g(t_0,t)&\alpha^g(t_0,t)).
    \label{bog_dynamics}
\end{equation}

From the equation describing the time evolution of the modes in terms of the initial conditions and from the definition of $\mathfrak{J}(t)$ we directly obtain that
\begin{equation}
    \mathfrak{B}(t_0,t)=\mathfrak{J}(t)T(t_0,t)\mathfrak{J}^{-1}(t_0),
\end{equation}
with the $\beta$-coefficients given by
\begin{equation}
\beta^f(t_0,t)=\frac{e^{-i\kappa(t)}}{2\omega_0\omega_{\perp}}
\left[ e^{i\Theta^2(t)}\Delta^2(t)\Delta^1(t_0) -e^{-i\Theta^1(t)}\Delta^1(t)\Delta^2(t_0) \right], \label{betacoeffs}
\end{equation}
where
\begin{equation} \label{Deltapl}
    \Delta^l(t) = \omega_{\perp}f_1(t)-(-1)^l\Omega^l(t)f_2(t)
\end{equation}
and recall that $\omega_{\perp}$ and $\Omega^l(t)$ are given in \eqref{omegaperp} and in \eqref{Omegal}, respectively.

Similarly, $\beta^g(t_0,t)$ satisfies an equation analogue to \eqref{betacoeffs} except for a global minus sign, and with $f_i$ replaced by $g_i$. Due to this symmetry and to the relations between the functions $f_i$ and $g_i$, from now on we will focus on the study of $\beta^f(t_0,t)$. The results obtained can also be applied in a straightforward manner to $\beta^g(t_0,t)$.

As discussed in section \ref{sec_unitary}, the condition for the $\beta$-coefficients to be associated with a unitarily implementable Bogoliubov transformation is that they be Hilbert-Schmidt for each fixed finite time $t$. This condition translates into square integrability (with respect to $\textbf{p}$) at all times, i.e., that
\begin{equation}
    \int d^3\bfp\ |\beta^f(t_0,t)|^2 =\int_0^{2\pi} d\phi \int_0^{\pi} d\theta \ \sin{\theta}   \int_0^{\infty} dp \ p^2|\beta^f(t_0,t)|^2  
\end{equation}
be finite for each $t$. Provided that $|\beta^f(t_0,t)|^2$ does not diverge in the angular coordinates $(\theta,\phi)$, as their domains are bounded, the integrability of $|\beta^f(t_0,t)|^2$ will be satisfied if and only if the integral in $p$ converges for almost all fixed directions. In particular, as we are considering a massive Dirac field, the $\beta$-coefficients \eqref{betacoeffs} do not diverge in the infrared. Thus, we are only interested in the UV regime, where the integrability condition is satisfied if and only if $\beta^f(t_0,t)= \order{p^{-\alpha}}$,
for some $\alpha>3/2$ at all times $t$.

The next step is to analyze the UV behaviour of $\beta^f(t_0,t)$ in equation \eqref{betacoeffs} for a fixed angular direction. First, we note that
\begin{equation}
    \omega_{\perp}=\sqrt{p^2\sin^2{\theta}+m^2},
\end{equation}
where $\theta \in [0,\pi]$ is the polar angle between $\textbf{p}$ and $\textbf{A}$. For $\theta \in (0,\pi)$ we have $\omega_{\perp}=\order{p}$, while for $\theta=0,\pi$ we have $\omega_{\perp}=m=\order{1}$. As the subspace of $\mathbb{R}^3$ which corresponds to $\theta=0,\pi$ (i.e., the $z$-axis) has zero measure, we will consider $\theta\in (0,\pi)$ from now on. Taking also into account that $\omega=\order{p}$, it is direct to see that the prefactor of \eqref{betacoeffs} is $\order{p^{-2}}$ for $\theta\in (0,\pi)$. 

Consequently, the integrability of the $\beta$-coefficients is assured if and only if the expression in brackets in \eqref{betacoeffs} is $\order{p^{-\alpha+2}}$ for all $t$ and in almost all the angular directions. In addition, we will neglect all the possibilities of cancelling the first term of \eqref{betacoeffs} with the second one. The reason is that $\Delta^l$ (and, therefore, the complex structure $\mathfrak{J}$) would need to absorb the leading order terms of the dynamical solutions $\Theta^l$. This would imply a trivialization of the Bogoliubov transformation \eqref{bog_dynamics} (see section \ref{sec_classicalbog}), thus allowing the unitary implementation of the dynamics trivially. Then, we will assume the independence of both terms in \eqref{betacoeffs} throughout time evolution, so that this integrability condition becomes
\begin{align}
    \Delta^2(t)\Delta^1(t_0)=\order{p^{-\alpha+2}}, \quad \Delta^1(t)\Delta^2(t_0)=\order{p^{-\alpha+2}},
    \label{integrabilitycondition}
\end{align}
with $\alpha>3/2$ for each finite value of $t$ and for almost all directions.

The relation \eqref{idf1f2.a} forces $f_1(t)$ and $f_2(t)$ to converge for each finite value of time $t$ to a complex number whose module is less than or equal to $1$ for arbitrarily large values of \bfp, i.e., the UV regime. It can be easily seen that if that limit is $0$ for $f_1(t)$ or $f_2(t)$, then both $\Delta^1(t)$ and $\Delta^2(t)$ behave as $\order{p}$ and the condition \eqref{integrabilitycondition} is not satisfied for any $\alpha > 3/2$. Therefore, the dynamics cannot be unitarily implemented.

In the most general case where both $f_1(t)$ and $f_2(t)$ are $\order{1}$ and none of them converge to 0, thus $\Delta^1$ and $\Delta^2$ given by \eqref{Omegal} and \eqref{Deltapl} would
be $\order{p}$. Consequently, we need to cancel the leading orders of $\Delta^s$, for $s=1$ or $s=2$, so that the $\beta$-coefficients have the right UV behaviour. Before we do so, it is useful to write the ratio of $f_1$ to $f_2$ as
\begin{equation} 
    \frac{f_1}{f_2}=u^s, \quad u^s= (-1)^s\frac{\Omega^s}{\omega_{\perp}}+h
    \label{f1/f2}.
\end{equation}
This parametrization, using \eqref{idf1f2.a}, can be rewritten as 
\begin{equation}
    f_1=\frac{u^s}{\sqrt{|u^s|^2+1}}e^{i\varphi}, \hspace{0.5cm} f_2=\frac{1}{\sqrt{|u^s|^2+1}}e^{i\varphi},
    \label{f1f2phase}
\end{equation}
where $\varphi=\varphi(t)\in \mathbb{R}$ is an arbitrary phase. Let us stress that, at this point, $u^s$ and $h$ are arbitrary. However, by requiring that the leading orders of $\Delta^s$ vanish for some $s$, we find in \eqref{f1/f2} that $h$ must be a function of order $\order{p^{-\sigma}}$, for some $\sigma \geqslant 0$, in order to have $\Delta^s=\order{p^{-\sigma+1}}$. For the other function $\Delta^{l\neq s}$ no cancellation occurs and $\Delta^{l\neq s}=\order{p}$. As a result, $\Delta^2(t)\Delta^1(t_0)$ and $\Delta^1(t)\Delta^2(t_0)$ are $\order{p^{-\sigma+2}}$. This satisfies condition \eqref{integrabilitycondition} if and only if $\sigma > 3/2$. In particular, this shows that the first two leading orders of $\Omega^s/\omega_{\perp}$ must cancel. 

The coefficient $\beta^f(t_0,t)$ does not diverge on the angles as can be checked by substituting  \eqref{f1f2phase} into \eqref{betacoeffs}, provided that $h$ is chosen with a smooth dependence on $\theta$. It is important to emphasize that, in order to translate the symmetries of the equations of motion to the complex structure, $h$ should not depend on the angular coordinate $\phi$. Considering all of the requirements exposed above, the integrability of the $\beta$-coefficients is assured.

In summary, the temporal evolution can be (non-trivially) unitarily implemented if and only if neither $f_1$ nor $f_2$ vanish in the UV and \eqref{f1f2phase} is verified for $s=1$ or $s=2$ in the UV regime, where
\begin{equation}
    h=\order{p^{-\sigma}}, \quad \sigma>3/2,
    \label{orderh}
\end{equation} 
is a function that depends smoothly on $\theta$. It is important to remember that we assumed that the electromagnetic potential verifies the mild conditions discussed in section \ref{sec_asymptotic} so that \eqref{formtheta} holds.

The requirements of the preservation of the symmetries of the equations of motion and the unitary implementation of the dynamics have reduced the possible complex structures to a unique family characterized by the integrability of the $\beta$-coefficients associated to the time evolution Bogoliubov transformation. As the density of the number of created particles and antiparticles throughout the evolution of the vacuum is given by
\begin{equation}
    N(t)=\frac{1}{2}\int_{\mathbb{R}^3} d^3\textbf{p} \ \left(|\beta^f(t,t_0)|^2+|\beta^g(t,t_0)|^2\right),
\end{equation}
this family of complex structures is equivalently characterized by a well-defined number of created particles at finite times. 

However, it is essential to emphasize that there is still some freedom in the selection of the particular complex structure within the family, encoded in the choice of $h$. This translates into an ambiguity in the particular value of $N(t)$ at finite times, which means that we need more criteria in order to characterize it completely. In particular, recent studies in hybrid loop quantum cosmology have already proposed additional requirements so that the Dirac Hamiltonian has nice mathematical properties. They further reduce the ambiguity by requiring for example that the fermionic backreaction is finite and that the Hamiltonian is diagonal asymptotically \cite{PhysRevD.98.063535,PhysRevD.99.063535,Elizaga_Navascu_s_2019}.

Let us note that when there is no electromagnetic field, the general quantization scheme followed here is unitarily equivalent to the usual Minkowski quantization. Indeed, in this case with $A=0$, the anisotropy introduced when working with $\gamma_0\gamma_3$ eigenvectors is irrelevant when compared to an isotropic treatment, since both descriptions are related through a unitary transformation which does not mix positive and negative frequencies.

\subsection{Unitary implementation of the dynamics for different potentials}

Let us now prove that a complex structure defined by $(f'_1,f'_2)$ which unitarily implements the dynamics for an electromagnetic potential $A'_{\mu}$ cannot unitarily implement the temporal evolution for a different potential $A_{\mu}$. In fact, if that happened we would have
\begin{equation}
    \frac{f'_1}{f'_2}=(-1)^s\frac{\Omega^s}{\omega_{\perp}}+h',
\end{equation}
with $\Omega^s$ satisfying \eqref{Omegal} for the potential $A_{\mu}$ (with $s=s'$; for $s\neq s'$ the result will be the same) and $h'=\mathcal{O}(p^{-\sigma'})$, $\sigma'>3/2$. Since $(f_1,f_2)$ verify \eqref{f1/f2}, it would imply that
\begin{equation}
    \frac{f_1}{f_2}-\frac{f'_1}{f'_2}=\order{p^{-\gamma}},
\end{equation}
with $\gamma>3/2$. However, this cannot hold, as the leading order of this difference of quotients can be easily seen to be
\begin{equation}
    \frac{f_1}{f_2}-\frac{f'_1}{f'_2}=\frac{q(A-A')}{\omega_{\perp}}[1+(-1)^s\cos{\theta}]+\cdots,
\end{equation}
which is $\mathcal{O}(p^{-1})$ for $\theta\in (0,\pi)$ unless $A_{\mu}=A'_{\mu}$. Note that we fixed in section \ref{section_fermionic} the temporal gauge $A_{\mu}(t_0)=0$. Had we not done so, then we would have obtained that for $A_{\mu}'=A_{\mu}+\text{constant}$ the quantizations should be equivalent.

In particular, this result states that the Minkowski quantization, obtained when $A_{\mu}=0$, does not unitarily implement the temporal evolution as long as there is a non-vanishing electromagnetic field. 

\subsection{Uniqueness of the quantization} \label{sec_uniqueness}

To which extent the requirements of preservation of the symmetries and unitary implementation of the temporal evolution reduce the ambiguity in the selection of the complex structure? Particularly, we will study if the quantum representations which admit a unitary implementation of the dynamics (characterized in the previous section) are unitarily equivalent. 

With this objective in mind, let $(a,b^*)$ and $(\tilde{a},\tilde{b}^*)$ be two sets of time-dependent annihilation and creation variables which allow for a unitary implementation of the dynamics. They satisfy relations of the form \eqref{abest}, being $\mathfrak{J}$ and $\tilde{\mathfrak{J}}$, respectively, the corresponding matrices defining their complex structures. It is not difficult to see that both sets of variables are related by a Bogoliubov transformation given by
\begin{equation}
    \mqty(a\\b^*)=\mathfrak{H} \mqty(\tilde{a}\\\tilde{b}^*),\hspace{0.5cm}  \quad\mathfrak{H}=\mathfrak{J}\tilde{\mathfrak{J}}^{-1}=
     \mqty(\kappa^f&\lambda^f\\\lambda^g&\kappa^g). \label{JJ-1}
\end{equation}
Using again Shale's theorem \cite{Shale:1962,RUIJSENAARS1978105}, both Fock representations $(a,b^*)$ and $(\tilde{a},\tilde{b}^*)$ will be unitarily equivalent if and only if $|\lambda^f|^2$ and $|\lambda^g|^2$ are integrable with respect to $\textbf{p}$. It is direct to see from \eqref{relationsbea} and \eqref{JJ-1} that
\begin{equation}
    |\lambda^f|=|\lambda^g|=|f_1 \tilde{f_2} - f_2\tilde{f_1}|,
    \label{lambda}
\end{equation}
where we followed the notation introduced in \eqref{abest}, with a tilde for the components of $\tilde{\mathfrak{J}}$. By hypothesis, $f_i$ and $\tilde{f}_i$ satisfy \eqref{f1/f2} in the UV regime, for some $s,\tilde{s}=1,2$ and $h=\order{p^{-\alpha}}$, $\tilde{h}=\order{p^{-\tilde{\alpha}}}$, with $\alpha,\tilde{\alpha} >3/2$. Substituting these relations into \eqref{lambda} we obtain
\begin{equation}
    |\lambda^f|=|\lambda^g|= |f_2\tilde{f}_2  ( u^s-\tilde{u}^{\tilde{s}} ) |.
\end{equation}
On the one hand, if $s=\tilde{s}$, then $|\lambda^f|=|\lambda^g|=\order{p^{-\min\{\alpha,\tilde{\alpha}\}}}$ and both representations are unitarily equivalent. On the other hand, if $s\neq \tilde{s}$, then $|\lambda^f|=|\lambda^g|=\order{1}$ (for $\theta\in(0,\pi)$), so both representations would not be unitarily equivalent. This is due to the fact that the relative sign between the functions $f_1$ and $f_2$ is different from the relative sign between $\tilde{f}_1$ and $\tilde{f}_2$. However, according to \eqref{relationsbea}, if we exchanged $\tilde{f}_i$ and $\tilde{g}_i$, then the relative sign between $\tilde{g}_1$ and $\tilde{g}_2$ would be the same as the one between $f_1$ and $f_2$. According to \eqref{abest}, this exchange between $\tilde{f}_i$ and $\tilde{g}_i$ is equivalent to a change in the convention of what we define as particles and antiparticles. Therefore, we can interpret that the inequivalence between these two representations is due exclusively to artificially choosing two different conventions ($s\neq \tilde{s}$) for the concepts of particle and antiparticle. Analogous results also appear in the study of the uniqueness of the Fock quantization of free Dirac fields in non-stationary curved spacetimes \cite{Cortez:2020rla}.

\section{Conclusions} \label{sec_conclusions}

In the study of a massive charged fermionic field coupled to a spatially homogeneous electric field, we have dealt with the reduction of the ambiguities in the process of canonical quantization. In particular, we consider that the physically relevant complex structures should preserve the symmetries of the system (the translational invariance due to the homogeneity of the external field and the decoupling between the modes in the equations of motion). In addition, we also impose that they should allow the unitary implementation of the dynamics with two main objectives: assuring throughout the evolution of the vacuum both the physical equivalence of the quantizations and the finiteness of the number of created particles at finite times.

This work aims to translate the analysis carried out in non-stationary curved spacetimes for free fields \cite{Cortez:2019orm,Cortez:2020rla} to the Schwinger effect. This was first done for a charged scalar field on \cite{Garay2020}. Other approaches with similar purposes are also found on the literature \cite{Dabrowski_2016}. However, our approach allows to obtain a characterization of the Fock representations assuring the compatibility with the requirements listed above at all finite times and not only asymptotically. Another strong point of the formalism followed here is the generality of the electromagnetic fields for which it applies. This family of external fields should satisfy certain mild time-dependence conditions and includes as a particular case those vanishing asymptotically.  

The main result of our analysis is that the physically reasonable requirements imposed on the quantization succeed in restricting the ambiguities to one unique equivalence class, which has been characterized. More precisely, when asking the complex structure to preserve the symmetries of the classical system and unitarily implement the dynamics, the freedom in its selection is reduced to just a choice of a function (for each mode) which has to decay sufficiently fast in the UV regime. The infinite possibilities for the selection of this function generates a family of unitarily equivalent complex structures characterized by a well-defined number of created particles at finite times. Thus, in this work we do not propose a unique candidate of this observable, but a selection of unitarily equivalent ones. Additional theoretical requirements, possibly based on experimental data, might help on reducing even more this residual ambiguity. This issue has already been studied in recent works for Dirac fields in cosmology, demanding the Hamiltonian to have nice physical and mathematical properties \cite{PhysRevD.98.063535,PhysRevD.99.063535,Elizaga_Navascu_s_2019}. This issue will be addressed elsewhere.

The choice of a privileged gauge is a direct consequence of the homogeneity of the electric field coupled to the matter field so that the spatial translational invariance is preserved in the quantization. In the case in which the electric field had spatial inhomogeneities, other procedures may be taken into account. The inclusion of perturbative electromagnetic inhomogeneities in the external field, including magnetic components, would provide a more realistic analysis of the Schwinger effect, allowing to establish a connection between theory and potential experiments. This issue has been recently analyzed numerically using the Dirac-Heisenberg-Wigner formalism and the Furry-picture quantization in \cite{PhysRevD.101.096003,PhysRevD.101.096009} and will be of central interest in our future works.

We have also shown that complex structures satisfying these criteria explicitly depend on the specific external electromagnetic field. As a consequence, a quantization based on the Minkowski vacuum (i.e., the corresponding to a vanishing electromagnetic field) does not unitarily implement the dynamics throughout the evolution of the vacuum in the presence of an electromagnetic field, in agreement with \cite{Ruijsencharged}. On the other hand, we have shown that the usual Minkowski vacuum is equivalent to a vacuum from our distinguished equivalence class in the case of a vanishing electric field. Consequently, in the frequently analyzed case in the literature of electromagnetic fields vanishing for $t\rightarrow \pm \infty$, the asymptotic behaviour of the complex structure (and then of the number of created particles) is the same both for the usual Minkowski quantization and for every quantization from our distinguished equivalence class. 

Finally, some other approaches to the Schwinger effect, such as the quantum kinetic approach \cite{Smolyansky1997DynamicalDO,Fedotov_2011}, usually make use of a particular quantization which diagonalizes the Dirac Hamiltonian. We leave a thorough comparison between other approaches and our proposal for the future. 

\acknowledgments

The authors are grateful to G. Garc\'ia-Moreno for useful discussions. This work has been supported by Project. No.
MICINN FIS2017-86497-C2-2-P from Spain (with extension Project. No. MICINN PID2020-118159GB-C44 under
evaluation).
AAD acknowledges financial
support from IPARCOS through ``Ayudas para la realizaci\'on de
Trabajos Fin de M\'aster del
Instituto de F\'isica de Part\'iculas y del Cosmos''.

\bibliographystyle{JHEP}
\bibliography{Schwinger}

\end{document}